\documentclass[10pt, journal]{IEEEtran}
\usepackage[colorlinks,linkcolor=red,anchorcolor=blue,citecolor=blue]{hyperref}
\usepackage{textcomp}
\usepackage{tikz}
\usepackage{color} 
\definecolor{circle}{RGB}{47, 135, 151} 
\usepackage{pifont}

\usepackage{graphicx}
\usepackage{url}
\usetikzlibrary{arrows, positioning, calc}
\tikzstyle{vertex}=[draw,fill=black!15,circle,minimum size=18pt,inner sep=0pt]
\usepackage{multicol}
\usepackage{multirow}
\usepackage[utf8]{inputenc}
\usepackage{authblk}
\usepackage[FIGBOTCAP]{subfigure}
\usetikzlibrary{shapes.geometric}
\usepackage{adjustbox}
\usepackage{bm}
\usepackage[utf8]{inputenc}
\usepackage{amsthm}
\usepackage[english]{babel}
\usepackage{bm}
\usepackage{amssymb}
\usepackage{float}
\usepackage{flushend}

\usepackage{cite}
\usepackage{amsmath,amssymb,amsfonts}
\usepackage{algorithm} 
\usepackage{algorithmicx} 
\usepackage[noend]{algpseudocode}
\algnewcommand{\algorithmicand}{\textbf{ and }}
\algnewcommand{\algorithmicor}{\textbf{ or }}
\algnewcommand{\OR}{\algorithmicor}
\algnewcommand{\AND}{\algorithmicand}

\usepackage{booktabs}
\usepackage{adjustbox}
\usepackage{colortbl} 
 
\usepackage[flushleft]{threeparttable} 
\usepackage{algorithm} 
\usepackage{algorithmicx} 

\begin{document}
\bstctlcite{IEEEexampleBSTcontrol}

\title{In-Network Market Prediction Using Machine Learning and Limit Order Books}

\author{Xinpeng~Hong\href{https://orcid.org/0000-0001-8525-6424}{\protect\includegraphics[scale=0.1]{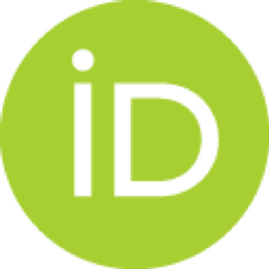}}, Changgang~Zheng\href{https://orcid.org/0000-0003-1894-722X}{\protect\includegraphics[scale=0.1]{orcid/icon.pdf}}, Joshua~Lilley\href{https://orcid.org/0009-0003-9449-9075}{\protect\includegraphics[scale=0.1]{orcid/icon.pdf}}, Stefan~Zohren\href{https://orcid.org/0000-0002-3392-0394}{\protect\includegraphics[scale=0.1]{orcid/icon.pdf}}, and~Noa~Zilberman\href{https://orcid.org/0000-0002-3655-2873}{\protect\includegraphics[scale=0.1]{orcid/icon.pdf}}
\thanks{Xinpeng Hong, Changgang Zheng, Joshua Lilley, Stefan Zohren, and Noa Zilberman are with the Department of Engineering Science, University of Oxford, United Kingdom.}
}

\maketitle

\begin{abstract}
Machine learning is significantly transforming algorithmic trading, yet the requirement for rapid execution speeds persists. While both aspects aim to boost profitability, embedding advanced machine-learning techniques with reduced trading latency presents a notable challenge. Adopting in-network machine learning, which involves offloading inference to programmable network devices, offers a delicate equilibrium in this trade-off. In this paper, we present LOBIN, a solution that utilizes machine learning within the network for market prediction based on high-frequency market data feeds. LOBIN is adept at constructing limit order books and performing inference directly within programmable switches. When compared to server-based benchmarks, LOBIN not only predicts future stock price movements with higher throughput but also maintains robust machine learning performance. It achieves over a 10\% reduction in latency compared to the NASDAQ order-matching server benchmark and delivers microsecond-level latency. Furthermore, the machine learning performance of LOBIN can be further enhanced through the adoption of a hybrid deployment approach that integrates both the switch and the servers. Our evaluation demonstrates that among all data feeds of evaluated stocks, the application of hybrid deployment results in approximately 45\% of the traffic and 38\% of the total potential transaction value being processed within switches without server intervention, reducing latency while ensuring that the average change in error rate of predictions remains at around 3\% relative to benchmarks based solely on server use.
\end{abstract}


\IEEEpeerreviewmaketitle

\section{Introduction}\label{ch1-Introduction}

Algorithmic trading has been growing over the past few decades as financial firms automate processes traditionally done by human traders. It uses computer algorithms to automatically execute orders under preset trading instructions~\cite{hendershott2011does}. As an essential form of algorithmic trading, high-frequency trading (HFT) is characterized by placing larger numbers of orders within a minimum time and being able to react quickly under changing market conditions~\cite{gomber2015high}. The internal latency of HFT systems that do not incorporate ML has been measured in microseconds~\cite{baldauf2020high}.

The rise of artificial intelligence (AI) further drives the growth of algorithmic HFT, with machine learning (ML) approaches becoming widespread in this field~\cite{kearns2013machine, huang2019automated}. However, the increasing complexity of ML models used also challenges existing trading systems, driving the need for a decrease in the delays in processing data, running models, and executing trades based on the insights gained. Furthermore, this also results in an escalating demand for more CPU cycles and electrical power within the HFT industry~\cite{wired}, which are considered detrimental to long-term and sustainable finance.

In-network computing offloads applications to run within programmable network devices~\cite{ tokusashi2019case}. In-network ML, as a specific type of in-network computing, deploys pre-trained ML models within network devices and conducts inference there for lower latency, higher throughput, and more efficient power utilization~\cite{sanvito2018can, xiong2019switches}. By design, in-network ML provides a practical solution for lowering latency in time-sensitive financial applications, such as those in trading scenarios, while also promoting AI and finance sustainability and curbing environmental impact through lower power use.

\begin{figure}[t]
	\centering
	\includegraphics[width=1\columnwidth]{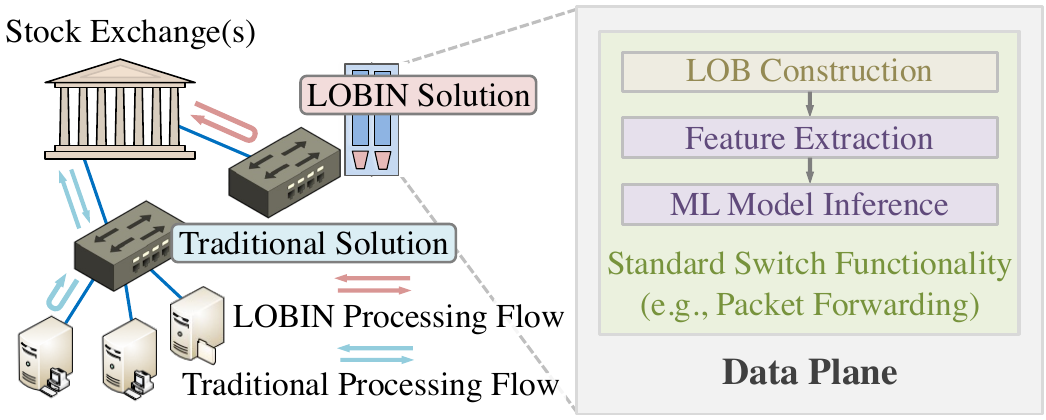}
	\vspace{-1.5em}
	\caption{General working scenario of LOBIN.}
	\label{fig:overall}
	\vspace{-1.7em}
\end{figure}

With limited prior works on in-network ML for time-critical financial uses, this paper provides an in-depth study on the application of in-network ML to a common HFT problem: predicting future price movements from market microstructure signals. This strategy has been proven to be feasible and profitable~\cite{kearns2013machine}. There are a number of previous works focusing on market forecasting utilizing electronic limit order books (LOBs) combined with ML models~\cite{kercheval2015modelling, ntakaris2018benchmark, zhang2019deeplob, tsantekidis2020using}. For a particular stock, a real-time LOB is constructed from unmatched limit orders that are predetermined with specific prices. It contains a wealth of information that can be used as ML features~\cite{baruch2005benefits}.

In this paper, we design and implement LOBIN (Limit Order Books In Network), an in-network prototype for future price movement forecasting by maintaining a LOB based on market-by-order (MBO) data feeds. Figure~\ref{fig:overall} shows a general working scenario of LOBIN compared with traditional solutions. The deployment location of LOBIN enables immediate prediction and decision-making, effectively circumventing the delay caused by traditional processing flow and server-based inference. Deployed on a Tofino~\cite{tofino} hardware switch, Tofino~2 (emulation environment), and a BMv2~\cite{bmv2} software switch model, LOBIN demonstrates the feasibility of the solution. According to the evaluation results, LOBIN can achieve a minimal loss of prediction performance compared to server-based implementation while lowering latency. By using a hybrid deployment strategy, LOBIN can achieve even higher ML performance while still retaining its ultra-low-latency advantage.

The main contributions of this paper are as follows:
\begin{itemize} 
    \item We study the application of in-network ML to market prediction from microstructure data and provide a proof of concept demonstrating its feasibility. This is the first of its kind to explore time-sensitive financial applications of in-network ML. 
    \item We design and implement a prototype that builds and updates LOBs in the programmable data plane based on high-frequency market data feeds, entailing the development of a complex data structure. We integrate the workflow of building and updating LOBs with ML-related processes, deploying it on both hardware and software programmable network devices. 
    \item We evaluate the prototype within a local networked-system testbed in terms of both ML prediction performance and networking performance. We compare different ML models, stock sections, and implementation solutions. While maintaining ML-based functionalities, the prototype achieves microsecond-level latency, reducing it by over 10\% compared to the server benchmark.
    \item We realize further improvements using a hybrid deployment approach across stocks, in terms of the accuracy, error rate, fraction of ML inference decisions handled by the switch, and latency, as a function of the on-switch ML classification confidence threshold. The analysis shows that hybrid deployment is both viable and beneficial, improving prediction performance, maintaining ultra-low latency, and reducing load on servers.
\end{itemize}


\section{Background and Motivation}\label{ch2-Background}

This section provides background about basic concepts and related works from three different domains: ML-based market prediction, trading acceleration using network devices, and in-network ML. It also motivates the need to explore the intersection of ML, programmable networking, and trading.

\subsection{ML for Market Prediction}

Centering on buying and selling assets in the marketplace, financial trading is an area where the application of ML approaches has become mainstream. Since financial time series data are inherently non-stationary and nonlinear, containing high noise~\cite{cheng2015time}, the applicability of traditional statistical methods is often constrained when dealing with this type of data. Aiming for profitability, ML models have been proven to have the capability to help with almost every point in the trading process, including forecasting price movement, generating trading signals, optimizing order execution, and making trading decisions~\cite{huang2019automated}. 

Among all aforementioned tasks, predictions on stock market price movement remain a big challenge because future fluctuations are influenced by countless internal and external factors, such as economic factors, investor sentiment, company performance, and industry performance~\cite{obthong2020survey}. To perform better in financial forecasting, researchers have applied and assessed different ML approaches based on historical data~\cite{ballings2015evaluating, patel2015predicting, basak2019predicting}. 

ML model complexity is continuously growing, using massive financial data more efficiently, laying an increasingly high burden on traditional processor-based platforms, and leading to a decrease in computational speed. However, in the scope of HFT, intense competition among traders requires lightning-speed real-time trade execution. Low latency is significantly crucial for generating profits because many of them are often based on short-lived opportunities such as cross-market arbitrage and breaking news~\cite{huang2019automated}. In today's competition, a small fraction of trading firms with the fastest HFT systems continue to amass a large share of trading revenues~\cite{baron2019risk}. Therefore, accelerating ML-based market prediction remains a challenge.

\subsection{Network Devices for Trading Acceleration}

To achieve lower latency, researchers focus not only on the optimization of trading algorithms and strategies but also on system-level solutions based on software design and hardware implementation. To date, different network devices have been used for trading acceleration, including application-specific integrated circuits (ASICs), graphic processing units (GPUs), or field-programmable gate arrays (FPGAs)~\cite{nurvitadhi2016accelerating}. Some studies used FPGAs for accelerating market data feed processing or trading applications, e.g.\ ~\cite{morris2009fpga, tang2016scalable} while others also utilized customized network interface cards (NICs) and optimized software for lower latency~\cite{subramoni2010streaming}. Significant progress was driven by the industry in developing hardware-based acceleration that can support financial services applications, e.g.\ \cite{vmware, amd}. However, none of the previous works has attempted to accelerate ML-based market prediction \textit{for making trading decisions within the network itself}. Given the innovative benefits that in-network computing and in-network ML can provide in terms of latency reduction, they may become a potential solution for this gap. If proven to be beneficial, a new generation of trading systems may be derived in the future. Since ASICs can offer superior processing performance compared to their counterparts~\cite{lockwood2012low}, this work uses switching ASICs for the lowest end-to-end latency.

\subsection{In-Network ML}

As the root of in-network computing, network programmability has facilitated network evolution. The emergence of software-defined networking (SDN) enabled networks to be intelligently controlled by software applications~\cite{scott2013sdn}. A specialized language, Programming Protocol-independent Packet Processors (P4), was developed for configuring how network devices process and forward packets within the data plane~\cite{bosshart2014p4}. Protocol-Independent Switching Architecture (PISA) is a common architecture that represents the data plane model in P4 based on a programmable match-action pipeline~\cite{pisa}.

Some P4 targets, such as the behavioral model (BMv2) switch and P4 platform on Raspberry Pi (P4Pi), use standard CPU to run packet forwarding programs~\cite{laki2021p4pi}, while some others are based on hardware including FPGAs~\cite{ibanez2019p4}, switches (e.g., Intel Tofino), and NICs (e.g., NVIDIA BlueField~2). P4 programs run on these target devices for packet forwarding and computing, offering programmability within the data plane. This provides a seedbed for offloading server applications to programmable network devices, which drove the emergence of in-network computing.

In-network computing offloads applications in part, or in full, to the data plane. It leverages lower power overheads, as well as higher process efficiency of network devices~\cite{tokusashi2019case}. In-network computing has been applied in several areas, including caching~\cite{jin2017netcache}, DNS~\cite{tokusashi2019case}, and distributed systems~\cite{dang2020p4xos}. Some studies focused on benefiting ML applications by offloading some parts of ML functions into the data plane such as feature extraction and weight aggregation~\cite{sapio2019scaling, lao2021atp}. To date, a number of ML models have been implemented for in-network ML, making their application to different fields possible~\cite{zheng2024planter}. However, the range of in-network ML use cases is still limited and needs further extension. Most relevant works focused on fields such as security and anomaly detection~\cite{lee2020switchtree, zang2023toward}, while the domain of financial trading was barely explored. In an early work, we considered market prediction, but only using stateless MBO messages~\cite{zheng2022automating}. Our experiments show that ML models perform poorly when using only raw fields from MBO messages as features, akin to a random classifier. This limitation motivates the use of LOBs as model inputs, as they offer richer information for better modeling~\cite{nagy20238}. However, incorporating complex structures like LOBs within the data plane poses significant implementation challenges.

\section{Overview of LOBIN}\label{ch3-LOBIN}

This section provides an introduction to MBO data and LOBs, along with an explanation of how LOBs update with market data feeds. Our solution, LOBIN, is subsequently introduced, with its key ideas and technical design explained.

\subsection{Limit Order Books}

Market-by-order (MBO) is an order-based, unencrypted data feed that provides the details of each trade instruction for a certain stock, directly from exchanges to traders~\cite{zhang2021deep}. It contains an order's timestamp, unique identifier, action (whether to add a new order, cancel an existing order, or update the price or quantity for the existing order), side (whether to buy or sell the given security), price, and quantity. Implicitly derived from MBO data, LOBs present a collection of unmatched limit orders that are waiting to be executed at pre-specified or better price levels~\cite{gould2013limit}. A limit order used to buy an asset at or below a pre-specified price is also called a bid order. In contrast, an ask order (or an offer) is to sell an asset at or above a given price level~\cite{zhang2019deeplob}. In a LOB, the two types of limit orders reside in the bid side and ask side, correspondingly. The mid-price is defined as the midpoint between the best (highest) bid price and the best (lowest) ask price.

In most of today's securities markets, bid orders and ask orders are executed following the price/time priority principle~\cite{parlour2008limit}. This principle dictates how limit orders are prioritized for execution: orders are operated firstly based on the best price, and if the specified prices of multiple orders are the same, the priority is given to the earliest one, breaking ties.

\begin{figure}[t]
	\centering
	\includegraphics[width=1\columnwidth]{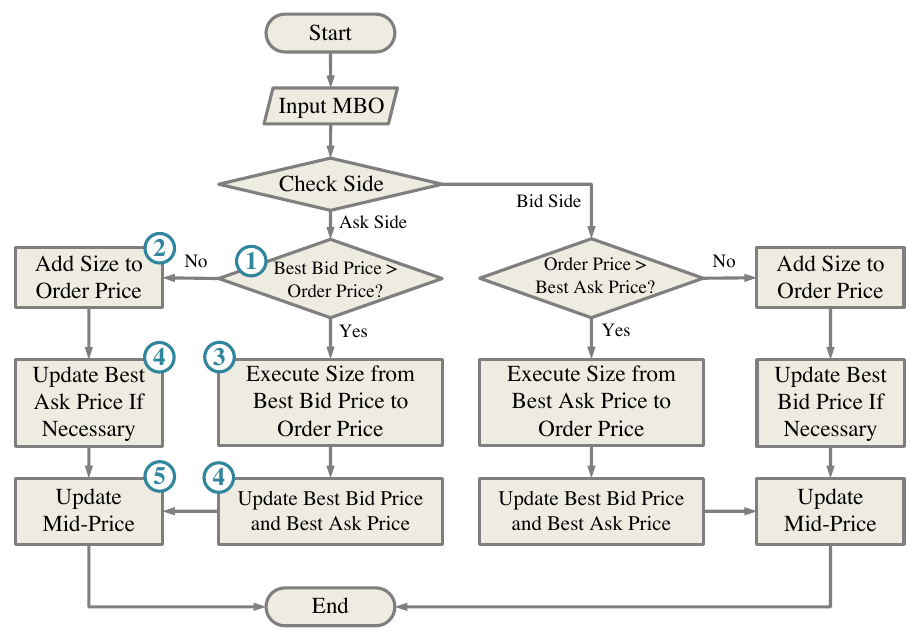}
	\vspace{-1.5em}
	\caption{Workflow of updating a LOB with MBO feeds.}
	\label{fig:workflow}
	\vspace{-1.5em}
\end{figure}

When executing a new order within the matching engine, the details it contains are used to decide how to update the LOB. Figure~\ref{fig:workflow} shows the workflow of updating a LOB with MBO data feeds. Taking an ask order as an example to illustrate the process, its limit price is compared to the best bid price, as shown in Step~$\textcircled{1}$. If the price of the new order is higher, the new order will be unmatched and reside at its price level of the LOB, as illustrated in Step~$\textcircled{2}$. If the price of the new order is equal to or lower than the bid, meaning that it crosses the bid-ask spread, as depicted in Step~$\textcircled{3}$, it will be matched with the unexecuted order(s) at the best bid price level. When any quantity of the new order remains after matching, it will be matched with the unexecuted order(s) at lower price level(s), as long as their price is bigger or equal to the price of the limit order. If unexecuted order(s) with higher price(s) than the new order has (have) been matched, the new order's remaining quantity will become an offer at its price level. As indicated in Step~$\textcircled{4}$, the best bid price and the best ask price within the LOB are updated if needed -- those changes result in the volatility of the mid-price, which is shown in Step~$\textcircled{5}$. Updating the LOB with a bid order is similar. 

\begin{figure*}[t]
	\centering
	\includegraphics[width=1\linewidth]{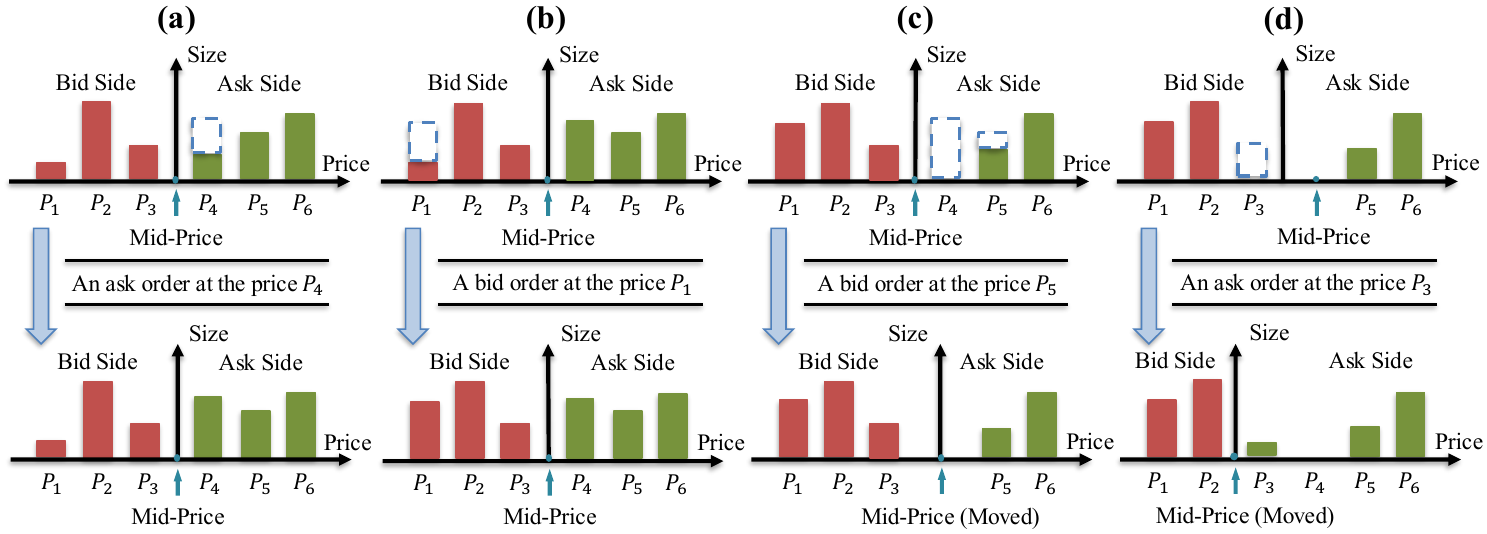}
	\vspace{-2em}
	\caption{An illustration of how MBO data updates a LOB.} 
	\label{fig:example}
	\vspace{-1.7em}
\end{figure*}

Figure~\ref{fig:example} illustrates how MBO data updates a LOB using four examples. Figure~\ref{fig:example}(a) presents how a new ask order with the price $P_4$ updates a LOB. Since $P_4$ is on the ask side, the quantity of the order settles with the existing volume at $P_4$. Similarly, Figure~\ref{fig:example}(b) shows that the size of a new bid order with the price $P_1$ resides at $P_1$ of the LOB. Figure~\ref{fig:example}(c) demonstrates the situation when a new bid order with a price higher than the best offer is used to update the LOB. In the example, the price of the bid order is $P_5$ while the existing best offer is $P_4$. The new order is matched with the existing orders at $P_4$ first, and then with the ones at $P_5$ if the unexecuted orders at $P_4$ are not sufficient for matching. After execution, the best offer becomes $P_5$, and thus the mid-price changes from the average of $P_3$ and $P_4$ to $P_4$. Last, in Figure~\ref{fig:example}(d), the LOB needs to be updated with a new ask order at the price $P_3$. Since the price of the new order equals the current best bid of the LOB, only unexecuted orders at $P_3$ are eligible for matching. If they are not enough to match the whole quantity of the new order, the remaining volume of it then resides at $P_3$ and $P_3$ becomes the best offer. The mid-price then changes to the average of $P_2$ and $P_3$ as the figure shows.

\subsection{LOBIN's Design}

LOBIN accelerates stock price movement prediction through the construction and updating of LOBs, a complex data structure, in the programmable data plane. For a given individual stock, when a new MBO message is received, LOBIN updates the LOB. It then extracts features, obtaining information on both price and quantity from different levels on the bid and ask sides of the LOB. The mid-price at each point in time is also used to create labels, representing the direction of price changes.

\begin{figure}[t]
	\centering
	\includegraphics[width=1\columnwidth]{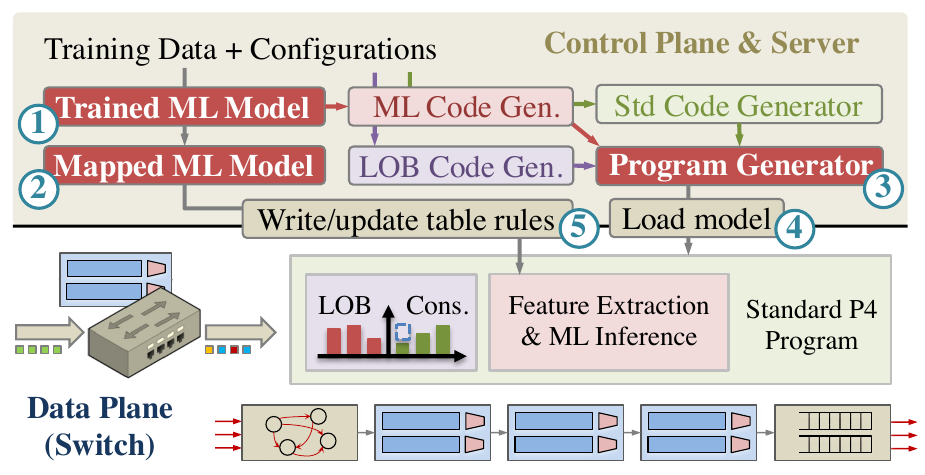}
	\vspace{-1.7em}
	\caption{System design of LOBIN.}
	\label{fig:systemdesign}
	\vspace{-1.65em}
\end{figure}

Figure~\ref{fig:systemdesign} shows the system architecture of LOBIN, including components on the server, control plane, and data plane. In the first step, running on a server, historical market data feeds are used to train an ML model, using LOB-based features. The trained model $\textcircled{1}$ is mapped to the data plane $\textcircled{2}$, and table entries are generated. The mapping includes generating a P4 program $\textcircled{3}$, which contains both the mapped model and LOB-related logic: LOB construction and update, and feature extraction. The generated P4 program is loaded to the programmable data plane $\textcircled{4}$, while table entries are loaded through the control plane $\textcircled{5}$. This design allows for direct prediction results in the data plane using trained ML models.

To further enhance LOBIN's performance, its design supports a hybrid deployment, combining a small model on the switch with a large model on the backend~\cite{zheng2024iisy}. In a hybrid deployment, a price-movement prediction in the switch will be labeled only if the confidence level of the prediction is high. Otherwise, the MBO would be sent to a server for prediction by a larger-size model. Such a deployment can reduce latency for most MBO messages and ML inferences, without compromising on ML prediction performance. 

\section{Implementation}\label{ch4-Implementation}

LOBIN is implemented using P4 on BMv2 and Intel Tofino switch-ASIC. Both implementations are based on the same concept: the LOB is initialized with an equal volume at each price level before transactions are executed. For each incoming MBO, the LOB is updated. After each message, the mid-price is computed based on the highest bid and the lowest offer at that timestamp, and LOB status features are extracted and used for ML inference. 

Programmable data planes are limited by different constraints, such as the amount of memory, number of stages, and permitted operations. This section explains using pseudocode the update process of a LOB within a data plane, and the solutions used by LOBIN to overcome the key challenges to enabling the logic within switch hardware.

\subsection{Software Switch Implementation}

BMv2 is an open-source behavioral software switch maintained by P4 workgroups~\cite{bmv2}, serving as a functional development target that can be adapted for various environments, such as P4Pi~\cite{laki2021p4pi} and eBPF~\cite{ebpf}. It is often used as a reference switch for functionality evaluation. As BMv2 is not as constrained as switch-ASICs, it is used as the first P4 data plane target for the general LOBIN process shown in Figure~\ref{fig:workflow}. Algorithm~\ref{alg:update1} presents the pseudocode of the updating process of a LOB with a bid order. The algorithm of updating a LOB with an ask order is similar and omitted for brevity. 

The algorithm uses three registers (\begin{math}R_{a}\end{math}, \begin{math}R_{b}\end{math}, and \begin{math}R_{l}\end{math}) to store the lowest ask price, the highest bid price, and current state of the LOB (the current quantity at each price level), respectively. These variables are stateful and need to be maintained and updated over time. Other variables in Algorithm~\ref{alg:update1} are user-defined metadata, used as intermediates for updating the LOB state and the best prices. First, the algorithm extracts the current best bid and ask prices (line 16). When the order price is lower than the best ask price, the algorithm increases the size at the order's price level and updates the best bid price if needed (lines 17-20). If the order price is higher, the algorithm iterates from the best ask price to the order price, fulfilling any matching orders and updating the best bid and ask prices as required (lines 21-33). Last, it records the updated best bid and ask prices back into registers (line 34).

\subsection{Tofino Implementation}

\setlength{\textfloatsep}{0.5cm}
\begin{algorithm}
	\caption{LOB update with a new bid order (General)} 
	\label{alg:update1}
	\begin{algorithmic}[1]
            \State - ${R}_{l}$: An array tracking LOB states (Register Array).
            \State - ${R}_{a}$: Position of lowest ask price (Register).
            \State - ${R}_{b}$: Position of highest bid price (Register).
            \State - $I$: Position of price levels in LOB. 
            \State - ${P}_{a}$: Position of lowest ask price (Metadata).
            \State - ${P}_{b}$: Position of highest bid price (Metadata).
            \State - ${P}_{o}$: Price level position of a new order.
            \State - ${S}_{o}$: Size (unexecuted) of a new order.
            \State
            \Function{UpdateMinAsk}{$I$} 
                \While {${R}_{l}[I] == 0$}
		          \State $I\leftarrow I+1$
		      \EndWhile
                \State ${P}_{a}\leftarrow I$ \Comment{\textcolor{gray}{Update the lowest ask price.}}
            \EndFunction
            \State
            \Function{Main}{${P}_{o}, {S}_{o}$}
            \State ${P}_{a}, {P}_{b}\leftarrow{R}_{a}, {R}_{b}$ \Comment{\textcolor{gray}{Read values from registers.}}
        \If {${P}_{o}<{P}_{a}$} \Comment{\textcolor{gray}{If order resides in LOB.}}
    	\State ${R}_{l}[{P}_{o}]\leftarrow{R}_{l}[{P}_{o}]+{S}_{o}$ \Comment{\textcolor{gray}{Update LOB.}}
    		\If {${P}_{o}>{P}_{b}$}
        		\State ${P}_{b}\leftarrow{P}_{o}$ \Comment{\textcolor{gray}{Update the highest bid price.}}
    		\EndIf
		\Else \Comment{\textcolor{gray}{Order needs to be matched.}}
		    \For {$I = {P}_{a}, \ldots, {P}_{o}$} \Comment{\textcolor{gray}{Iteration for matching.}}
		        \If {${R}_{l}[I]>={S}_{o}$} \Comment{\textcolor{gray}{LOB order size at the ${I}_{th}$ price level is sufficient to match.}}
    		        \State ${R}_{l}[I]\leftarrow{R}_{l}[I]-{S}_{o}$ \Comment{\textcolor{gray}{Update LOB.}}
    		        \State \Call{UpdateMinAsk}{$I$}
                        \State $break$
		        \Else \Comment{\textcolor{gray}{LOB order size at the ${I}_{th}$ price level is insufficient to match.}}
		              \State ${S}_{o}\leftarrow{S}_{o} - {R}_{l}[I]$ \Comment{\textcolor{gray}{New unexecuted size.}}
    		        \State ${R}_{l}[I]\leftarrow0$ \Comment{\textcolor{gray}{Update LOB.}}
		              \If {$I == {P}_{o}$} 
		                \State ${R}_{l}[I]\leftarrow{S}_{o}$ \Comment{\textcolor{gray}{Unmatched size of the new order resides in LOB.}}
		                \State ${P}_{b}\leftarrow I$ \Comment{\textcolor{gray}{Update the highest bid.}}
                          \State \Call{UpdateMinAsk}{$I+1$}
		              \EndIf
		        \EndIf
		    \EndFor
		\EndIf
		\State ${R}_{a}, {R}_{b}\leftarrow {P}_{a}, {P}_{b}$ \Comment{\textcolor{gray}{Write values back to registers.}}
        \EndFunction
	\end{algorithmic} 
\end{algorithm}
\setlength{\floatsep}{0.1cm}

\begin{algorithm}
    \caption{LOB update with a new bid order on Tofino}
    \label{alg:update2}
    \begin{algorithmic}[1]
    \State - $F$: Recirculation flag ($1$ if a recirculated message).
    \State - ${S}_{I}$: Order size at the ${I}^{th}$ price level in LOB.
    \State - $a, b, c$: Temporary variables for comparisons.
    \State - Other variable definitions are as in Algorithm~\ref{alg:update1}.
    \State 
    \Function{Operation}{$I$} 
        \State ${S}_{I}\leftarrow {R}_{l}[I]$ \Comment{\textcolor{gray}{Read order size at the $I^{th}$ price level.}} 
        \State $a\leftarrow I - {P}_{a}$ \Comment{\textcolor{gray}{Distance from lowest ask price.}} 
        \State $b\leftarrow {P}_{o} - I$ \Comment{\textcolor{gray}{Distance from new order's price.}} 
        \State $c\leftarrow {S}_{o} - {S}_{I}$ \Comment{\textcolor{gray}{Gap between new order's unexecuted size and LOB size at the ${I}^{th}$ price level.}} 
        \If{$a\geq0$\AND$b>0$\AND$c\geq0$} \Comment{\textcolor{gray}{Order size at the ${I}^{th}$ price level is insufficient to match new order.}} 
            \State ${S}_{I}, {S}_{o}\leftarrow 0, c$ \Comment{\textcolor{gray}{Update LOB and unmatched size.}}
        \ElsIf{$a\geq0$\AND$b==0$\AND$c>0$} \Comment{\textcolor{gray}{The unmatched size of the new order resides in LOB.}}
            \State ${S}_{I}, {S}_{o}\leftarrow c, 0$ \Comment{\textcolor{gray}{Update LOB and unmatched size.}}
            \State ${P}_{b}\leftarrow I$ \Comment{\textcolor{gray}{Update the highest bid price.}}
            \State ${P}_{a}\leftarrow I+1$ \Comment{\textcolor{gray}{Update the lowest ask price.}}
        \ElsIf{$a\geq0$\AND$b==0$\AND$c==0$} \Comment{\textcolor{gray}{Order size at the ${I}^{th}$ price level is just enough to match.}} 
            \State ${S}_{I}, {S}_{o}\leftarrow 0, 0$ \Comment{\textcolor{gray}{Update LOB and unmatched size.}}
            \State ${P}_{a}\leftarrow I+1$ \Comment{\textcolor{gray}{Update the lowest ask price.}}
        \ElsIf{$a\geq0$\AND$b\geq0$\AND$c<0$} \Comment{\textcolor{gray}{Order size at the ${I}^{th}$ price level is more than sufficient to match.}} 
            \State ${S}_{I}, {S}_{o}\leftarrow{S}_{I}-{S}_{o}, 0$ \Comment{\textcolor{gray}{Update LOB size and unmatched order size.}}
            \State ${P}_{a}\leftarrow I$ \Comment{\textcolor{gray}{Update the lowest ask price.}}
        \EndIf
    \EndFunction
    \State 
    \Function{Main}{${P}_{o}, {S}_{o}$}
    \If{${F}==0$} 
    \State ${P}_{a}, {P}_{b}\leftarrow{R}_{a}, {R}_{b}$ \Comment{\textcolor{gray}{Read values from registers.}}
    \If{${P}_{o}\leq{P}_{b}$} \Comment{\textcolor{gray}{If order resides in LOB.}} 
        \State ${R}_{l}[{P}_{o}]\leftarrow{R}_{l}[{P}_{o}]+{S}_{o}$ \Comment{\textcolor{gray}{Update LOB.}}
    \ElsIf{${P}_{b}<{P}_{o}<{P}_{a}$} \Comment{\textcolor{gray}{If order resides in LOB and the highest bid price needs updating.}} 
        \State ${R}_{l}[{P}_{o}]\leftarrow{R}_{l}[{P}_{o}]+{S}_{o}$ \Comment{\textcolor{gray}{Update LOB.}}
        \State ${P}_{b}\leftarrow{P}_{o}$ \Comment{\textcolor{gray}{Update the highest bid price.}}
    \ElsIf{${P}_{a}\leq{P}_{o}$} \Comment{\textcolor{gray}{Order matches.}} 
        \For {$I = 0, \ldots, {P}_{h}$} \Comment{\textcolor{gray}{Iteration for operation.}} 
        \If{${S}_{o}>0$} \Comment{\textcolor{gray}{Unexecuted size exists.}}
            \State \Call{Operation}{$I$}
        \EndIf
        \EndFor
    \EndIf
    \State $F\leftarrow1$ \Comment{\textcolor{gray}{Update recirculation flag.}} 
    \ElsIf{$F==1$} 
    \State ${R}_{a}, {R}_{b}\leftarrow {P}_{a}, {P}_{b}$ \Comment{\textcolor{gray}{Write values back to registers.}}
    \For {$I = 0, \ldots, {P}_{h}$} 
    \State ${R}_{l}[I]\leftarrow{S}_{I}$ \Comment{\textcolor{gray}{Write back updated LOB size.}} 
    \EndFor
    \EndIf
    \EndFunction
    \end{algorithmic} 
\end{algorithm}

Intel Tofino is a P4-based programmable switch ASIC, which utilizes the Tofino Native Architecture (TNA), an architecture that is very similar to the Portable Switch Architecture (PSA) used in BMv2. Tofino offers Tbps (Terabit-per-second) data rate, with microsecond-scale latency~\cite{bf6064t}, which can potentially be deployed at the network edge or access and provide LOB services. However, similar to other programmable switch-ASICs, it has resource constraints such as stage count and memory capacity~\cite{hauser2022survey}. While BMv2 is suitable for P4 prototypes, using Tofino indicates the viability of commodity off-the-shelf devices in a real-world trading environment.

The constraints of Tofino's ASIC design, mean that the general solution previously described needs to be adapted to the hardware. To overcome the limitations of the platform, the process of updating the LOB with MBO data should be modified. The main implementation challenges are:

\begin{itemize} 
    \item \textbf{Lack of loops:} Basic workflow requires loops for several purposes. For instance, to locate maximum bid, minimum offer, or when executing a new bid order that needs to be served by multiple offer levels in the LOB, due to its quantity. While loop unrolling is one solution, its cost in resources and stage consumption is high.
    \item \textbf{Cost of comparisons:} LOB construction and maintenance require a lot of price and order volume comparisons. For example, for an incoming offer, its price needs to be compared with the best bid and the best offer. If an offer is matched with a bid, their quantities need to be compared, and any remaining unmatched quantity needs to be handled. On hardware, each comparison consumes a processing stage and reduces scalability.
    \item \textbf{Registers access:} LOB must maintain state over MBO messages and be quickly updated, which requires the use of registers. However, a register can be accessed only once in the pipeline, and it is not allowed to read a register at the beginning of the pipeline and then write an updated value back at the end of the pipeline. To overcome this challenge, recirculation can be used, with one pass for reading and the second for writing.
\end{itemize}

In addition to the challenges above, the sequential call to \begin{math}if-else\end{math} conditions, as used in Algorithm~\ref{alg:update1}, can easily exceed the maximum number of pipeline stages. Consequently, LOBIN's processing flow is modified from a hierarchical structure into a flatter, parallel design. 

The key idea in the solution is using regular logic to update the quantity at a single price level in the LOB, and operating through all price levels using the same logic, regardless of the current order's price. The algorithm compares each price level with the price of the new bid order and the current best ask price (lines 8-9). It also compares the quantity of each price level with the order size to be matched (line 10). At each price level, the algorithm's decision-making is influenced by four possible scenarios arising from the comparisons' results: 1) The order size at the current price level falls short of fulfilling the order size to be matched (lines 11-12); 2) It is just enough to match the remaining order size (lines 17-19); 3) It surpasses the quantity needed for matching (lines 20-22); 4) The new order's remaining unmatched size can reside at the current price level (lines 13-16). Based on these outcomes, the algorithm decides if and how to operate on the current price level, as well as if to update the best bid and ask prices.

Algorithm~\ref{alg:update2} shows the pseudocode of updating a LOB with a bid order on Tofino. Updating an ask order follows a similar process.
While line \begin{math}22\end{math} of Algorithm~\ref{alg:update1} and line \begin{math}33\end{math} and \begin{math}39\end{math} of Algorithm~\ref{alg:update2} use \begin{math}for\end{math} commands, these are only for illustration purposes, and loop unrolling is applied in practice.

The P4 implementation of LOBIN integrates the LOB-based feature engineering and extraction code as a preprocessing step with ML inference code generated by Planter~\cite{zheng2024planter}, a framework for mapping trained ML models to programmable network devices. LOB generation and updates require 622 lines of P4 code for BMv2 and 680 lines for Tofino. Automatically generated inference code varies across different ML models. For instance, a decision tree model requires 390 lines for BMv2 and 165 lines for Tofino. The code for architecture and network functionality across both targets requires fewer than 100 lines. The implementation also supports Tofino~2 without further code changes. Moreover, the implementation supports a hybrid ML deployment.

\section{Evaluation}\label{ch5-Evaluation}

This section assesses LOBIN's performance using publicly accessible MBO data, focusing on ML performance, latency, and throughput. It includes a comparative analysis of LOBIN on commodity programmable switches and server-based benchmarks, employing different ML models. Additionally, LOBIN is evaluated combined with a hybrid deployment approach. The findings demonstrate LOBIN's improved throughput, reduced latency, and consistent prediction performance compared to server benchmarks.

\noindent\textbf{Dataset and Tools:} The dataset used in this study is NASDAQ’s Historical TotalView-ITCH sample data feeds~\cite{nasdaqnasdaq_site}. All MBO messages for each trading day are encapsulated within a compressed binary format file. The data format specification is publicly available~\cite{nasdaqnasdaq}. Due to the absence of data feeds spanning consecutive days in the data source, data from a single date is used. MBO messages were extracted on January 30, 2020, representing the most recent date available at the time of analysis. Three stocks are picked for the evaluation, representing companies in the NASDAQ Composite index with the largest market capitalization in three different sectors: Diamondback Energy Inc (NASDAQ: FANG, Sector: Energy), Exelon Corp (NASDAQ: EXC, Sector: Utilities), and AstraZeneca PLC (NASDAQ: AZN, Sector: Healthcare). Our evaluation prioritizes a comparative analysis of ML performance relative to established server benchmarks rather than focusing solely on absolute performance. With around 440K entries from three distinct stocks, the dataset provides a diverse representation of market behaviors, allowing for fair comparisons in ML performance. 
An open-source tool~\cite{zenodo} is used for reconstructing MBO messages of one specific stock from the binary format file. The Planter framework~\cite{zheng2024planter} is used for in-network ML deployment.

\noindent\textbf{Experimental Setup:} An APS-Networks BF6064T-X Intel Tofino switch is used for evaluation. The 64$\times$100G ports switch runs SDE (software development environment) 9.4.0. The system supporting Tofino~2 uses SDE 9.9.0 running on a server. Server-based experiments, including traffic generation, are conducted with two ASUS ESC4000A-E10 servers, equipped with AMD EPYC 7302P CPU and 256GB DDR4 RAM, using Ubuntu 20.04 LTS. Both servers are connected to the switch using NVIDIA ConnectX-5 100G NICs and direct attach cables. Scikit-learn (Sklearn) library~\cite{pedregosa2011scikit} is used for ML training and inference on the server.

\subsection{Prediction Performance}

Model prediction performance is tested for several different ML algorithms that are commonly used for forecasting future stock price movement, including k-means (KM), k-nearest neighbors (KNN), decision trees (DTs), random forests (RFs), and extreme gradient boosting (XGB)~\cite{ballings2015evaluating, patel2015predicting, basak2019predicting}. Some in-network ML models, like support vector machines and naive Bayes, are not feasible, as they consume more hardware stages, preventing their deployment together with LOB-related code.

ML model training uses up to ten price level volumes of a LOB and the mid-price as input features for server-based and BMv2-based prediction. For Tofino-based prediction, up to three price level volumes of a LOB and the mid-price are used, due to resource constraints on the switch. Server-based benchmarks are trained with unlimited-size models while avoiding overfitting, as evidenced by our experiments where server-based DTs reach up to 20 in depth and 100000 in leaf nodes, and both RFs and XGB include up to 300 trees, with similar maximum depth and leaf nodes. In contrast, switch-based models are of limited size. For example, switch-based DTs have a maximum depth of 4 and up to 1000 leaf nodes, while RFs and XGB are restricted to 4 trees, a maximum depth of 3, and a limit of 1000 leaf nodes. The labels are used to predict future mid-price movement (up, down, and stationary) over the next 100 ticks. A smoothing labeling method~\cite{ntakaris2018benchmark} is used to draw more consistent signals from highly stochastic financial feeds. Synthetic minority over-sampling technique (SMOTE) is applied to address class imbalance~\cite{chawla2002smote}. 

To explore performance gains with resources' scalability, the prediction performance of Tofino~2 is emulated. Tofino~2 is less resource-constrained than Tofino. As it has more stages, it is possible to use larger LOBs, extract more features, increase model sizes, or utilize ML models that consume more stages. 

\begin{table*}[htbp]\large	
\renewcommand\arraystretch{0.9}
\begin{adjustbox}{width=1\linewidth,center}
\centering
\begin{threeparttable}
\begin{tabular}{lcc>{\columncolor[gray]{0.93}}cc|cc>{\columncolor[gray]{0.93}}cc|cc>{\columncolor[gray]{0.93}}cc|cc>{\columncolor[gray]{0.93}}cc}
\toprule
\multicolumn{1}{l}{} & \multicolumn{16}{c}{\textbf{NASDAQ: FANG (Sector: Energy)}} \\ 
\midrule
\multicolumn{1}{l}{Target:} & \multicolumn{4}{c|}{Tofino} & \multicolumn{4}{c|}{Tofino 2} & \multicolumn{4}{c|}{BMv2} & \multicolumn{4}{c}{Server} \\ 
\midrule
\multicolumn{1}{l}{Model} & PRE & REC & F1 & ACC & PRE & REC & F1 & ACC & PRE & REC & F1 & ACC & PRE & REC & F1 & ACC \\ 
\midrule
KM & 34.91 & 33.75 & 13.43 & 23.81 & 30.85 & 33.91 & 14.05 & 24.20 & 23.68 & 33.32 & 27.59 & 52.90 & 23.68 & 33.32 & 27.59 & 52.90 \\
KNN & 39.89 & 33.59 & 12.93 & 23.13 & 40.94 & 33.65 & 13.00 & 23.20 & 41.66 & 36.63 & 34.51 & 44.01 & 57.76 & 60.72 & 58.01 & 64.58 \\
DT & 51.14 & 57.88 & 53.36 & 69.95 & 50.84 & 57.88 & 53.26 & 69.90 & 60.61 & 60.59 & 56.48 & 70.83 & 60.61 & 60.59 & 56.48 & 70.83 \\
RF & 62.46 & 58.27 & 54.78 & 69.46 & 51.28 & 57.70 & 53.29 & 69.93 & 52.03 & 59.26 & 54.60 & 70.93 & 59.20 & 61.49 & 56.29 & 70.72 \\
XGB & 51.14 & 57.88 & 53.36 & 69.95 & 50.84 & 57.88 & 53.26 & 69.90 & 67.65 & 59.69 & 56.84 & 71.92 & 59.82 & 61.90 & 58.17 & 68.79 \\
\midrule
\multicolumn{1}{l}{} & \multicolumn{16}{c}{\textbf{NASDAQ: EXC (Sector: Utilities)}} \\ 
\midrule
\multicolumn{1}{l}{Target:} & \multicolumn{4}{c|}{Tofino} & \multicolumn{4}{c|}{Tofino 2} & \multicolumn{4}{c|}{BMv2} & \multicolumn{4}{c}{Server} \\ 
\midrule
\multicolumn{1}{l}{Model} & PRE & REC & F1 & ACC & PRE & REC & F1 & ACC & PRE & REC & F1 & ACC & PRE & REC & F1 & ACC \\ 
\midrule
KM & 29.99 & 32.23 & 28.72 & 29.76 & 36.20 & 33.36 & 32.10 & 31.59 & 36.80 & 36.89 & 36.30 & 35.04 & 36.80 & 36.89 & 36.30 & 35.04 \\
KNN & 22.20 & 34.78 & 25.92 & 30.92 & 26.77 & 36.17 & 24.18 & 32.26 & 25.44 & 42.05 & 31.55 & 37.30 & 48.64 & 51.08 & 47.59 & 49.00 \\
DT & 50.84 & 44.87 & 43.37 & 48.03 & 49.50 & 45.04 & 43.92 & 48.21 & 55.36 & 59.81 & 53.38 & 55.32 & 60.12 & 62.21 & 59.18 & 59.41 \\
RF & 49.63 & 45.87 & 45.06 & 48.99 & 49.61 & 45.94 & 45.15 & 49.01 & 58.94 & 60.73 & 52.68 & 55.55 & 60.25 & 61.50 & 59.34 & 60.39 \\
XGB & 48.00 & 40.67 & 35.53 & 45.96 & 48.30 & 41.37 & 36.99 & 46.52 & 58.76 & 61.17 & 51.75 & 55.75 & 60.70 & 61.43 & 59.92 & 60.85 \\
\midrule
\multicolumn{1}{l}{} & \multicolumn{16}{c}{\textbf{NASDAQ: AZN (Sector: Healthcare)}} \\ 
\midrule
\multicolumn{1}{l}{Target:} & \multicolumn{4}{c|}{Tofino} & \multicolumn{4}{c|}{Tofino 2} & \multicolumn{4}{c|}{BMv2} & \multicolumn{4}{c}{Server} \\ 
\midrule
\multicolumn{1}{l}{Model} & PRE & REC & F1 & ACC & PRE & REC & F1 & ACC & PRE & REC & F1 & ACC & PRE & REC & F1 & ACC \\ 
\midrule
KM & 7.21 & 23.41 & 10.39 & 17.13 & 19.33 & 28.18 & 16.68 & 21.84 & 27.18 & 37.26 & 29.59 & 40.16 & 27.18 & 37.26 & 29.59 & 40.16 \\
KNN & 11.63 & 33.33 & 17.25 & 34.90 & 13.63 & 33.33 & 19.35 & 40.89 & 66.54 & 40.69 & 32.35 & 39.22 & 60.56 & 58.96 & 59.18 & 61.06 \\
DT & 29.60 & 39.09 & 33.67 & 44.93 & 29.54 & 39.01 & 33.61 & 44.84 & 50.55 & 50.21 & 46.92 & 54.33 & 57.14 & 52.43 & 49.85 & 55.17 \\
RF & 30.37 & 40.20 & 33.69 & 47.16 & 32.69 & 42.95 & 37.05 & 48.95 & 45.22 & 54.83 & 47.27 & 61.25 & 47.69 & 58.39 & 50.57 & 65.74 \\
XGB & 34.50 & 34.31 & 31.88 & 37.47 & 29.54 & 39.01 & 33.61 & 44.84 & 57.29 & 63.24 & 54.74 & 71.69 & 60.70 & 60.67 & 55.48 & 67.41 \\
\bottomrule
\end{tabular}
\end{threeparttable}
\end{adjustbox}
  \vspace{0.2em}
  \caption{ML prediction performance (\%) with three NASDAQ stocks from three different sectors. LOBIN runs limited-size models on BMv2 and Tofino (or emulated on Tofino~2). The benchmark runs on a server using Sklearn and unlimited-size models.}
  \label{ml}%
  \vspace{-2em}
\end{table*}%

Table~\ref{ml} presents experimental results using key ML metrics: macro precision (PRE), macro recall (REC), macro F1-score (F1), and accuracy (ACC). Precision measures the ratio of correct predictions in a class to all predictions for that class, and macro precision averages this across classes. Macro recall calculates the proportion of correct predictions per class against actual instances, averaged across all classes. The F1 score, a balance between precision and recall, is their harmonic mean. Accuracy reflects the overall correct predictions over total predictions. The F1 score is often preferred over accuracy as it better addresses class imbalance, providing a more comprehensive model evaluation, especially in skewed class distributions. All four metrics range from 0 to 1, with higher values indicating better performance.

The benchmarks' ML performance results, running on the server, achieve performance comparable to previous (ML-focused) studies that used similar features from LOBs for the same prediction task~\cite{zhang2019deeplob}. Their test results, using different neural networks, achieved F1-score ranging from 40.84\% to 76.58\%, on the same datasets and with the same prediction horizon of 100 as in our experiments. In comparison, the results of our ML benchmarks are on a par. Additional tests with recurrent neural networks (RNN) and long short-term memory (LSTM) networks across stocks yielded F1-scores of 70.92\% to 78.28\% after fine-tuning hyperparameters to validate the state-of-the-art performance achievable on servers.


As the results in Table~\ref{ml} show, the LOBIN solution's ML performance on a software switch (BMv2) aligns closely with server benchmarks across different models and stocks. Running on BMv2, it has an average accuracy loss of 4.39\% with an average F1-score loss of 6.47\%. Specifically, the precision and recall experience an average reduction of 3.54\% and 4.16\%, respectively. This difference is mainly due to the variation in model sizes that can be supported.

Notably, even on Tofino, LOBIN's performance loss is small for FANG stock, highlighting its strengths. For instance, across three tree-ensemble models, LOBIN's accuracy loss on Tofino is less than 1\%, with a loss of 3.15\% for the F1-score compared with server benchmarks. The F1-score loss is evidenced by a 3.32\% drop in average recall and a more substantial 4.96\% reduction in average precision. 

\begin{figure}[t]
	\centering	\includegraphics[width=1\columnwidth]{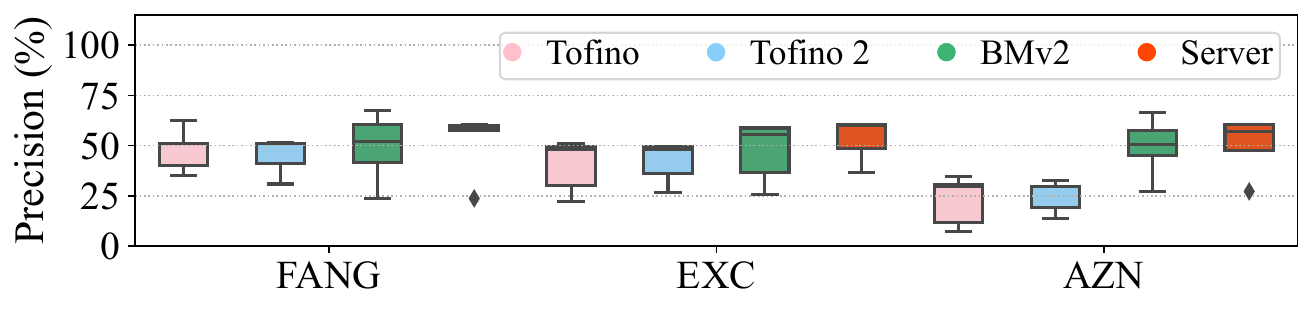}
	\vspace{0em}
 	\centering
\includegraphics[width=1\columnwidth]{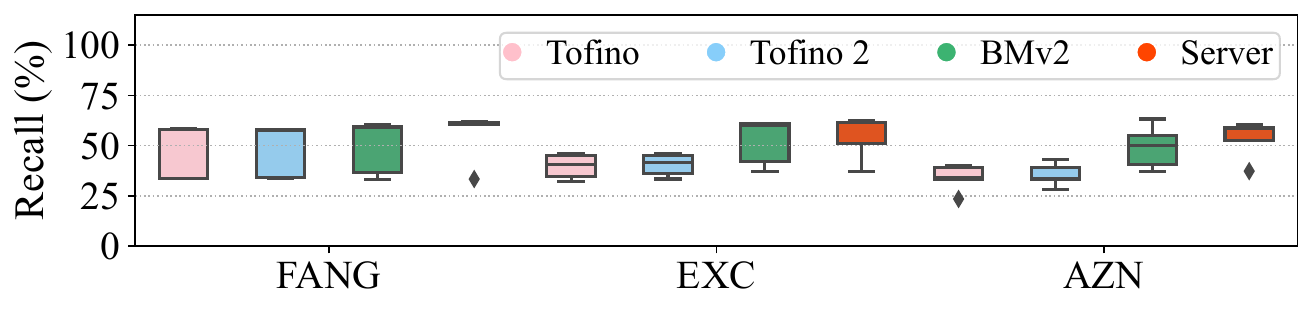}
	\vspace{0em}
 	\centering
\includegraphics[width=1\columnwidth]{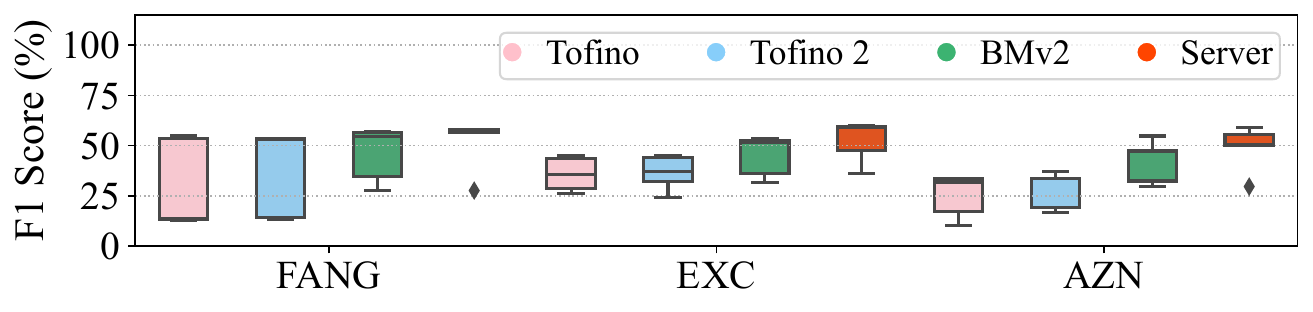}
	\vspace{-1em}
 	\centering
\includegraphics[width=1\columnwidth]{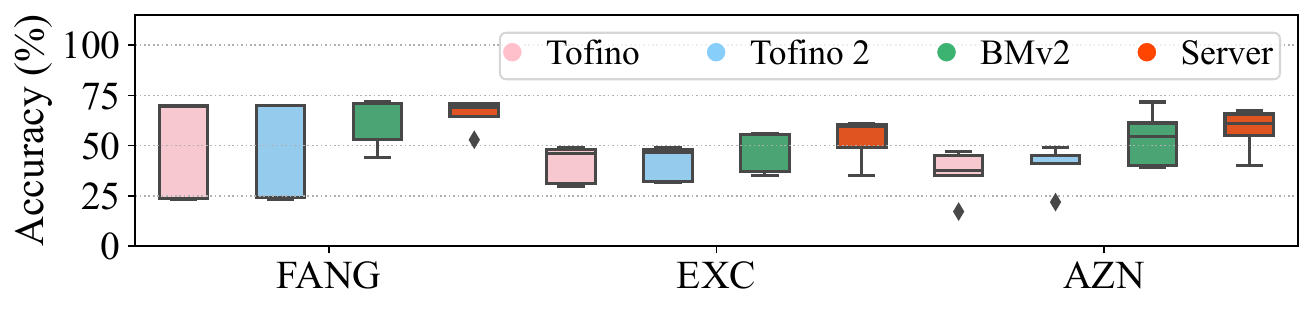}
	\vspace{-1em}
	\caption{Box plots of precision, recall, F1-score, and accuracy across ML models, for different targets, Tofino, Tofino~2, BMv2, and the server.}
	\label{box}
	\vspace{-1em}
\end{figure}

Overall, the ML performance of LOBIN, when deployed on Tofino, falls short of BMv2 and server-based solutions due to hardware resource limitations, restricting it to only supporting smaller-sized ML models and fewer features. These constraints also force LOBIN to construct smaller LOBs, thereby diminishing its sensitivity to mid-price shifts. Results from Tofino~2 show an improvement over Tofino as two additional features from a larger LOB can be supported. Consequently, LOBIN can achieve better ML performance on Tofino~2 hardware. Additionally, Tofino2 offers the flexibility to allocate its extra resources to support either larger ML models or larger LOBs with more features. Our experiments indicate that both approaches yield similar performances.

\begin{figure}[t]
	\centering	\includegraphics[width=1\columnwidth]{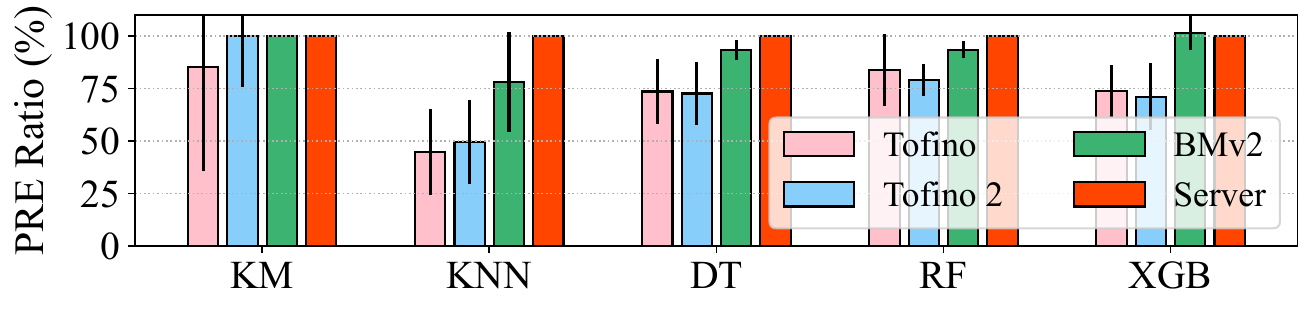}
	\vspace{0em}
 	\centering
\includegraphics[width=1\columnwidth]{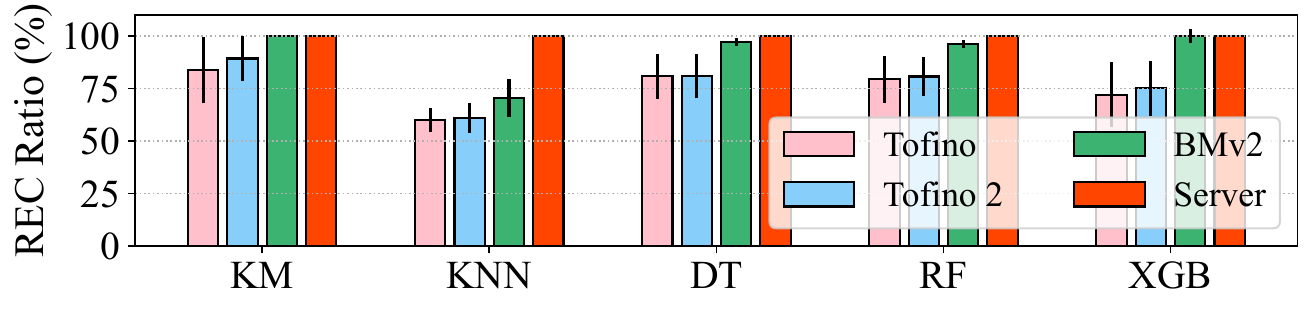}
	\vspace{0em}
 	\centering
\includegraphics[width=1\columnwidth]{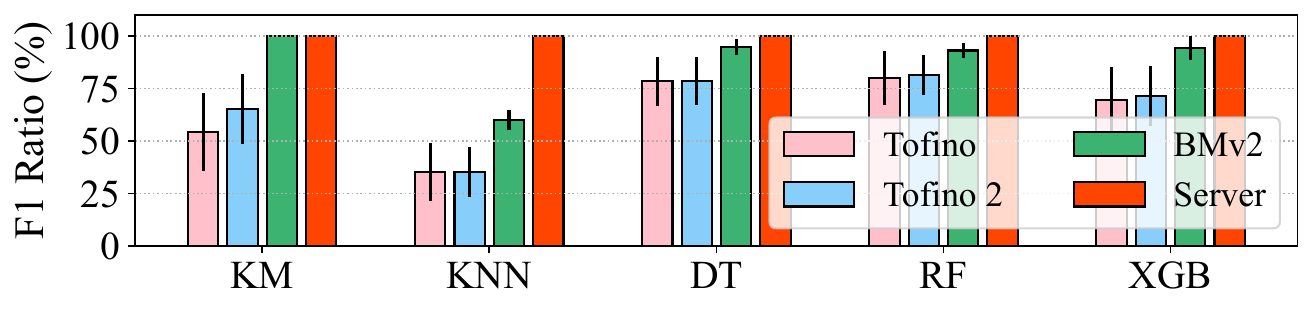}
	\vspace{-1em}
 	\centering
\includegraphics[width=1\columnwidth]{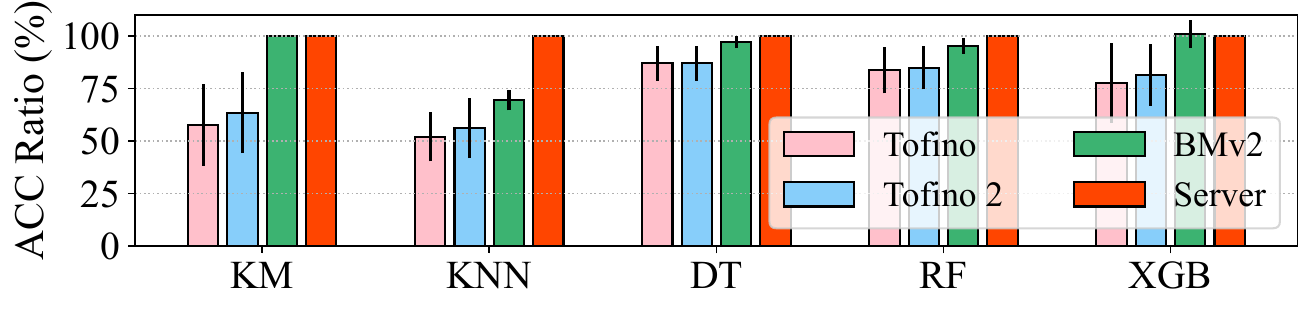}
	\vspace{-1em}
	\caption{The average ratio (\%) of precision, recall, F1-score, and accuracy across models relative to the benchmark, on Tofino, Tofino~2, and BMv2.}
	\label{ratio}
	\vspace{-1.3em}
\end{figure}

In further examining LOBIN's ML performance, the average ratios of all four metrics on different targets are computed, relative to server benchmarks. On average, Tofino attains 72.20\%, 75.21\%, 63.50\%, and 71.57\% of the ML benchmarks' precision, recall, F1-scores, and accuracy, respectively. The corresponding metrics for Tofino~2 fall within the range of roughly 66\% to 77\%. For BMv2, these ratios are notably higher, spanning from around 88\% to 93\%. Figure~\ref{box} shows box plots of all metrics on all targets across models, highlighting LOBIN's consistent pattern and robust performance for each stock, with performance differences across targets and benchmarks being most noticeable for AZN. A comparative analysis is also conducted to scrutinize the performance disparities among in-network models. The outcomes, delineated in Figure~\ref{ratio}, indicate that tree-based ensemble methods stand out by delivering low loss, high ratios, and consistent performance stability across stocks in ML predictions.

\subsection{Networking Performance}

The latency of Tofino is under NDA, therefore we report our measurements of pipeline and framework latencies relative to Tofino's reference switch (\textit{switch.p4}). LOBIN's relative pipeline latency is computed based on data reported by its SDE and includes one pipeline recirculation per MBO message. LOBIN's framework latency is measured between two servers, with a switch positioned in between. Precision Time Protocol (PTP), using the ptp4l toolkit, is used for the measurement, with timestamps taken on both servers. The relative framework latency is the ratio of the measured LOBIN latency to the measurement of simple forwarding through the switch. 

\begin{figure}[t]
	\centering
	\begin{minipage}{0.49\linewidth}
	\includegraphics[width=1\linewidth]{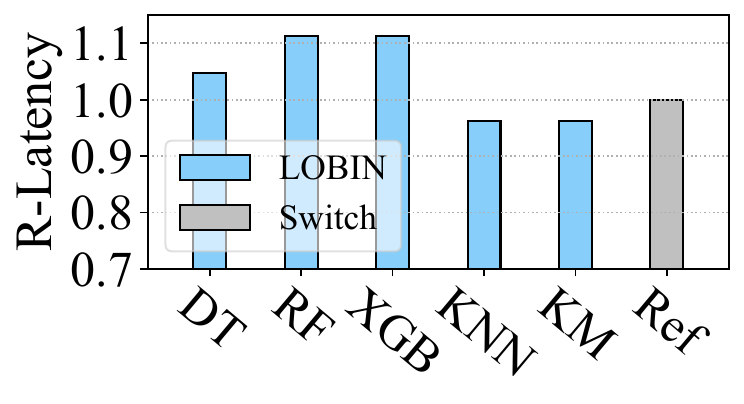}\vspace{-0.5em}\\\centering(a) Pipeline R-Latency
\end{minipage}
	\begin{minipage}{0.49\linewidth}
	\includegraphics[width=1\linewidth]{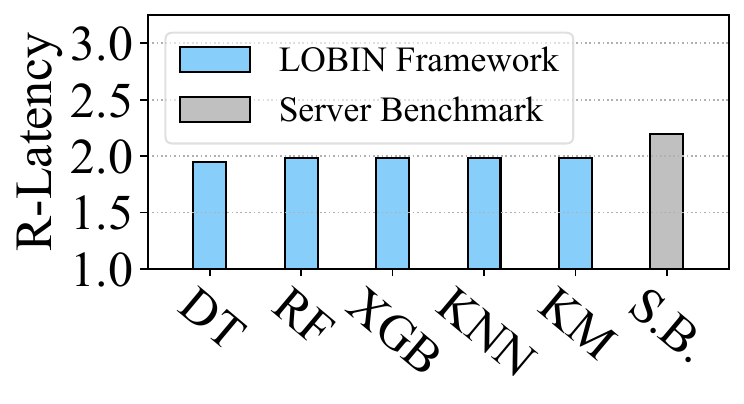}\vspace{-0.5em}\\\centering(b) Framework R-Latency
\end{minipage}
\vspace{-0.5em}
	\caption{(a) The pipeline relative latency (R-Latency) on Tofino for different models, measured for LOBIN with standalone ML and standalone $switch.p4$. (b) The framework R-Latency of LOBIN for different models, compared with a server benchmark (S.B.). (Stock: FANG).} 
	\label{fig:latency}
	\vspace{-1em}
\end{figure}

Measurements are performed across all three stocks, yielding consistent results with no significant discrepancies. Consequently, we use the FANG stock for illustrative purposes. As shown in Figure~\ref{fig:latency}~(a), all of LOBIN's models have a similar latency to the reference \textit{switch.p4}. This demonstrates that even under resource constraints, LOBIN itself achieves comparable latency to simple packet switching. On the framework level, as Figure~\ref{fig:latency}~(b) shows, the latency of LOBIN is higher than simple forwarding due to recirculation. However, LOBIN achieves its microsecond-level latency while doing on top of forwarding also LOB updates and ML inference. LOBIN also achieves over 10\% improvement compared to the NASDAQ order-matching server benchmark~\cite{bonart2017latency}, that only measures LOB updates without ML inference. This indicates even higher improvements if inference were included.

In a throughput test, packets are sent using DPDK 20.11.1 and PktGen 21.03.0. Two separate pipelines (32 ports) are connected in a snake configuration where each port is linked to the next one in the pipeline, with recirculation through unused pipelines. LOBIN achieves full 1.6Tbps ($16\times100Gbps$) on each of the switch pipelines, and 3.2Tbps in total.

\subsection{Hybrid Deployment Performance}

The prediction performance of LOBIN on Tofino can be further improved using a hybrid ML deployment~\cite{zheng2024iisy}. The performance of LOBIN with the implementation of a hybrid deployment strategy is evaluated across all three stocks. The baseline is an unlimited-size ensemble model running on the servers. On the switch, a limited-size model is deployed for inference. Messages with a low classification confidence level are subsequently forwarded to the servers for more in-depth classification. The switch is configured with a confidence level to decide the on-switch classification threshold.

\begin{figure}[t]
	\centering
	\begin{minipage}{0.49\linewidth}
	\includegraphics[width=1\linewidth]{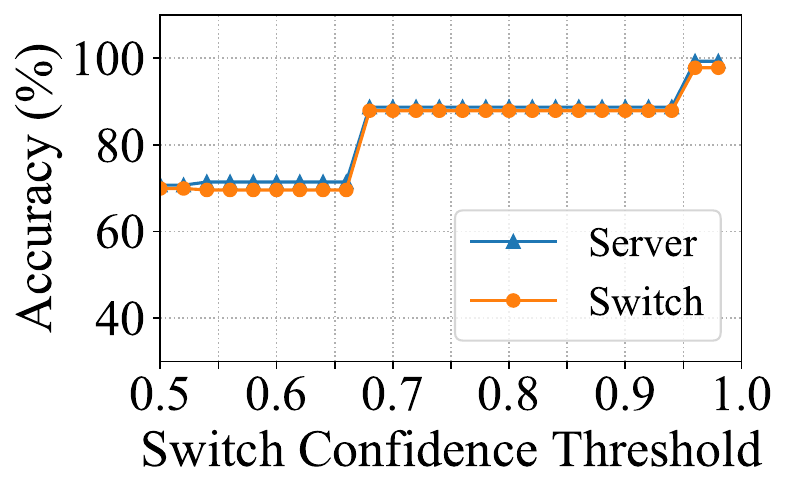}\vspace{-1em}\\
\end{minipage}
	\begin{minipage}{0.49\linewidth}
	\includegraphics[width=1\linewidth]{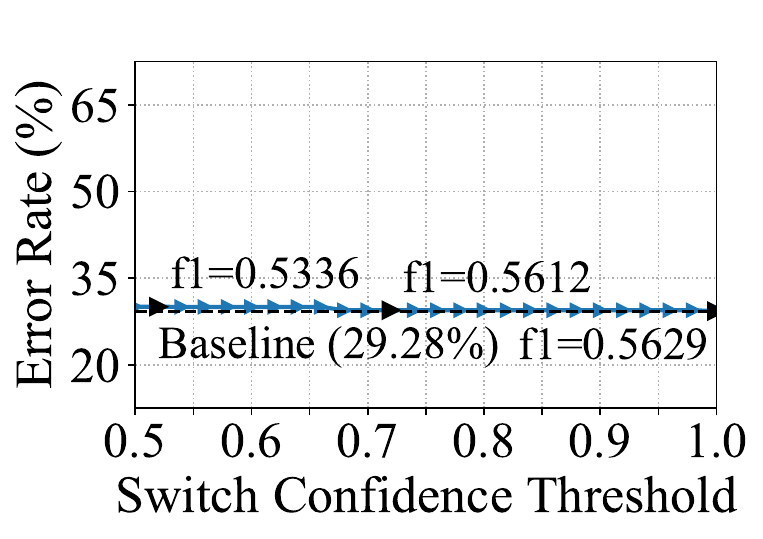}\vspace{-1em}\\
\end{minipage}
\centering
\begin{minipage}{0.49\linewidth}
	\includegraphics[width=1\linewidth]{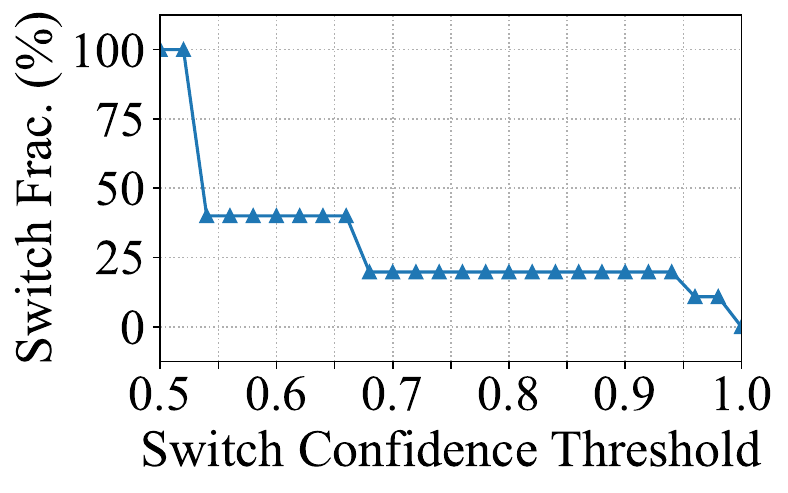}\vspace{-0.5em}\\
\end{minipage}
	\begin{minipage}{0.49\linewidth}
	\includegraphics[width=1\linewidth]{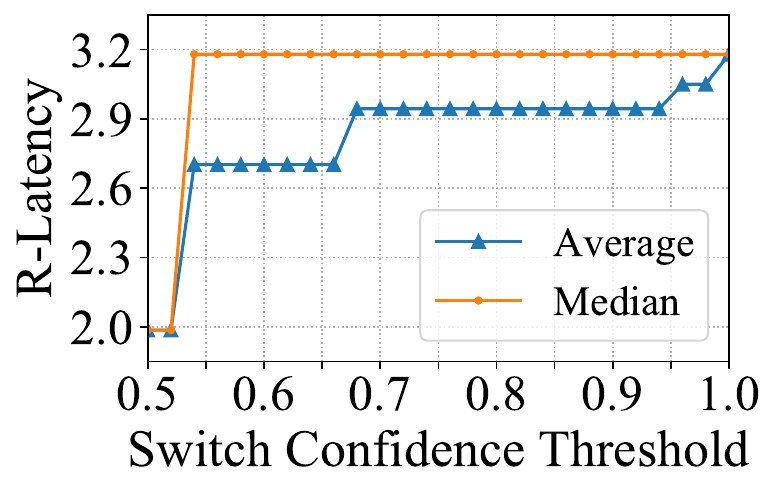}\vspace{-0.5em}\\
\end{minipage}
\vspace{-0.8em}\\\centering(a) Stock: FANG\\
\vspace{0.5em}
\centering
	\begin{minipage}{0.49\linewidth}
	\includegraphics[width=1\linewidth]{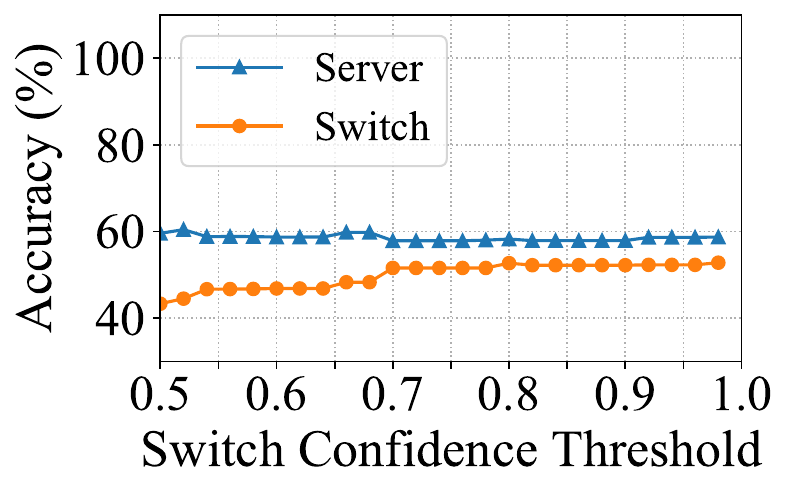}\vspace{-1em}\\
\end{minipage}
	\begin{minipage}{0.49\linewidth}
	\includegraphics[width=1\linewidth]{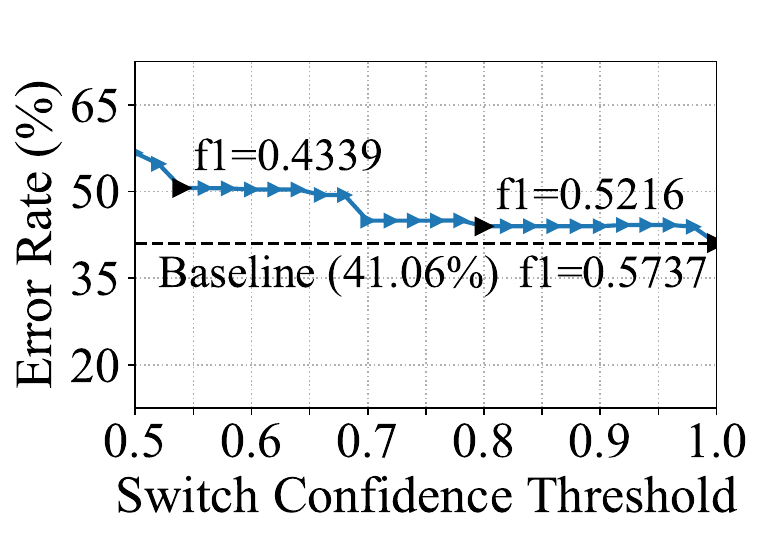}\vspace{-1em}\\
\end{minipage}
\centering
\begin{minipage}{0.49\linewidth}
	\includegraphics[width=1\linewidth]{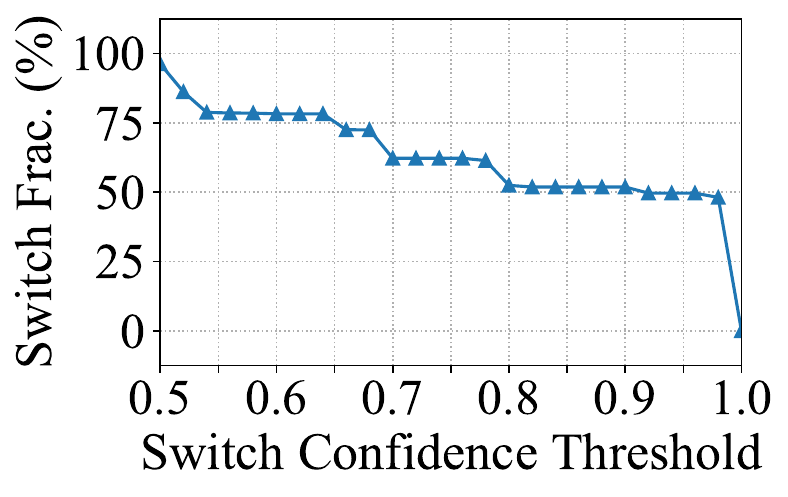}\vspace{-0.5em}\\
\end{minipage}
	\begin{minipage}{0.49\linewidth}
	\includegraphics[width=1\linewidth]{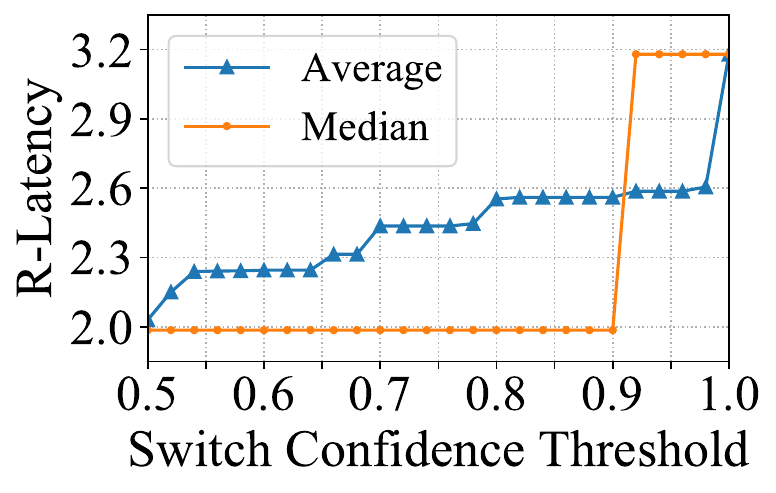}\vspace{-0.5em}\\
\end{minipage}
\vspace{-0.8em}\\\centering(b) Stock: EXC\\
\vspace{0.5em}
\centering
	\begin{minipage}{0.49\linewidth}
	\includegraphics[width=1\linewidth]{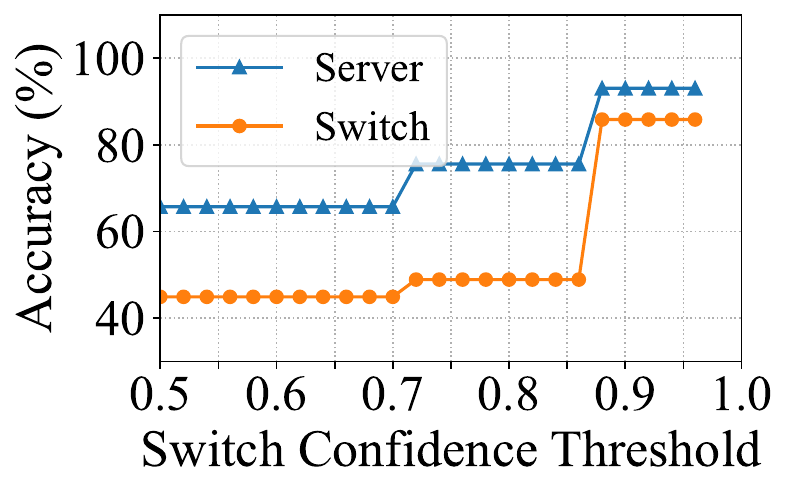}\vspace{-1em}\\
\end{minipage}
	\begin{minipage}{0.49\linewidth}
	\includegraphics[width=1\linewidth]{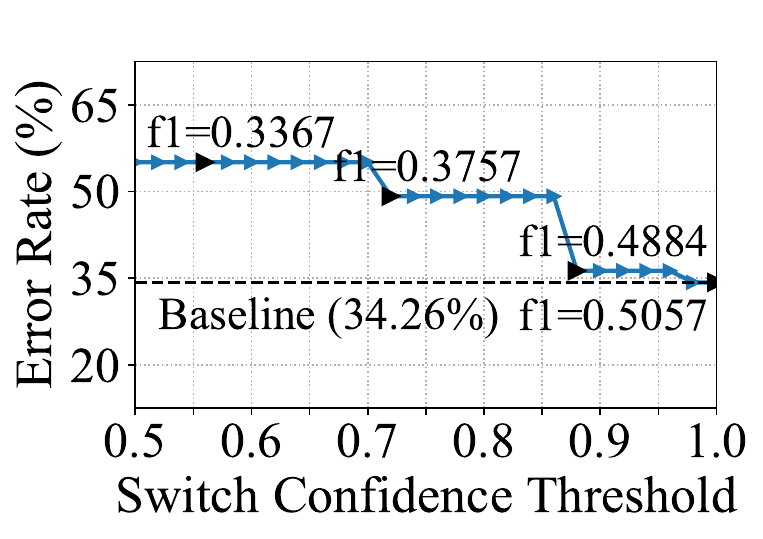}\vspace{-1em}\\
\end{minipage}
\centering
\begin{minipage}{0.49\linewidth}
	\includegraphics[width=1\linewidth]{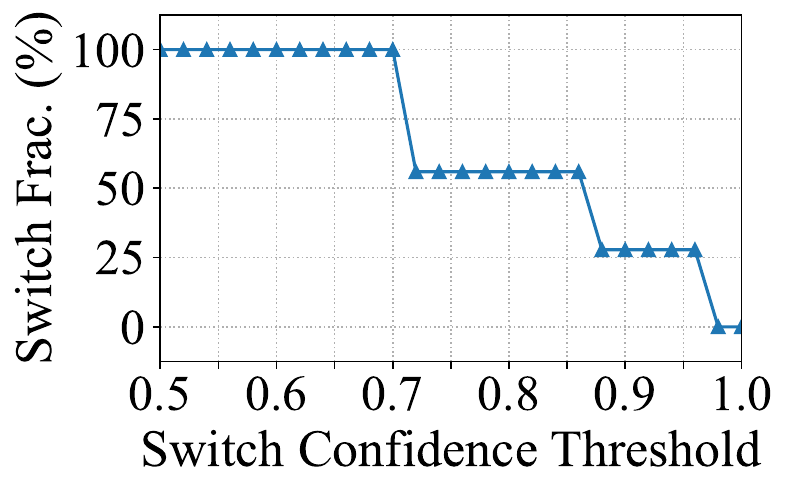}\vspace{-0.5em}\\
\end{minipage}
	\begin{minipage}{0.49\linewidth}
	\includegraphics[width=1\linewidth]{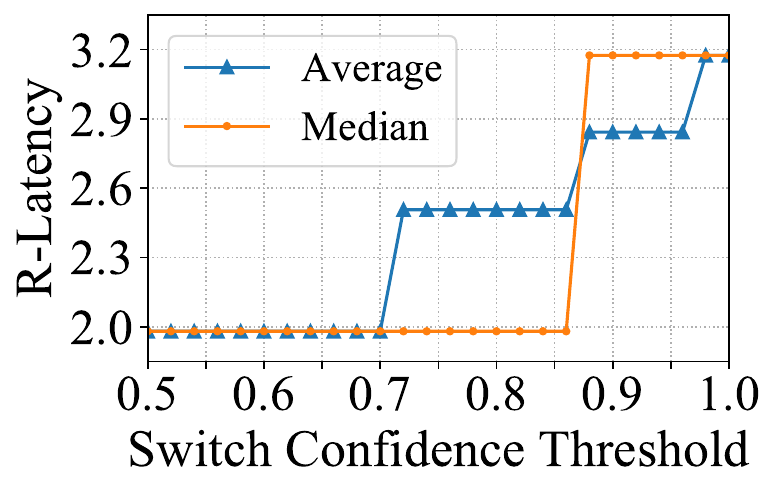}\vspace{-0.5em}\\
\end{minipage}
\vspace{-0.8em}\\\centering(c) Stock: AZN\\
\vspace{-0.5em}
	\caption{The accuracy, the error rate, the fraction (Frac.) of traffic offloaded by the switch, and the end-to-end average and median R-latency (latency relative to simple forwarding) of LOBIN with three stocks using Random Forest models, across switch classification confidence thresholds. ``Average" is calculated by dividing the total latency by the number of orders processed, whereas ``Median" refers to the latency of a single order that divides the feed into two halves, with one half having higher latencies and the other half lower.} 
	\label{hybrid}
	\vspace{-1.2em}
\end{figure}

Figure~\ref{hybrid} illustrates the relationship between the switch classification confidence threshold and four key performance metrics, including accuracy, misclassification rate, fraction of offloaded traffic, and framework latency relative to simple forwarding through the switch, when using a random forest model for classification with FANG, EXC, and AZN stocks. 

Several shared findings for the three stocks can be deduced: First, increasing the switch confidence threshold leads to enhanced prediction accuracy on the switch as well as the accuracy on the server for the instances served by the switch. As this threshold is raised, the overall rate of misclassification declines while the F1-score improves. The on-switch classification accuracy is lower compared to the on-server accuracy for the instances processed by the switch, because the model on the switch, which comprises only 4 trees, is significantly smaller than the one running on the server, which consists of 200 trees with unlimited depth and number of leaf nodes. Second, a higher confidence threshold on the switch results in more order feeds being sent to the server for processing. Consequently, this leads to a decrease in the proportion of traffic that the switch can offload. Third, sending more order feeds to the server for processing also leads to an increase in the average latency for each processed order. The sharp rise in median latency observed in all latency-related figures indicates the point at which half of the traffic is directed to the server, while the other half is managed entirely on switch. 

As Figure~\ref{hybrid}~(a) shows, for stocks like FANG, where LOBIN's on-switch classification accuracy is nearly equivalent to the server benchmark, offloading all inference to the switch results in a slight increase in error rate of less than 3\%. In such scenarios, the average latency can be reduced by approximately 9.5\%, compared with the NASDAQ server benchmark without using a hybrid setup. The hybrid deployment has nearly no impact on the error rate (within 0.2\% margin) while offloading approximately 20\% of the traffic to the switch.

For stocks such as EXC and AZN, where the ML performance gap between LOBIN's model on hardware and the server benchmark is relatively larger compared to stocks like FANG, adopting a hybrid deployment is more effective. Based on Figure~\ref{hybrid}~(b) and (c), by adjusting the switch confidence threshold to optimal levels, 0.98 for EXC and 0.96 for AZN, the increase in error rate can be maintained below 5\% for both. This approach enables direct on-switch processing of 48.12\% of EXC's orders and 27.81\% of AZN's orders. In practice, a suitable threshold can be selected for each stock to strike a balance across all metrics.

When considering the data feeds of all evaluated stocks collectively, a hybrid deployment leads to around 45\% of the traffic being processed directly by the switch, bypassing server intervention. In terms of transaction value, this equates to about 1.97 billion dollars out of a total of 5.13 billion, representing 38\% of the total value, being managed on-switch. This method also ensures that the average change in misclassification rate is maintained at around 3\%, highlighting its efficiency in large-scale trading feed processing.

\section{Discussion}\label{ch6-Discussion}

LOBIN contributes to latency reduction of short-term price predictions from LOBs \textit{By Design}. Even in comparison with inference solutions using SmartNICs or FPGA, LOBIN eliminates the latency of getting to the host, with minimal overhead beyond that of standard switch forwarding. While LOBIN is implemented on Tofino, using low-latency programmable switches can reduce latency further. LOBIN can also be deployed on multiple switches, each managing the LOBs of different stocks or distributing a large ML model~\cite{zheng2023dinc}, thus enhancing scalability.

While LOBIN is shown to be feasible on hardware targets, the resources restrict the increase in LOB size. LOBIN does not store MBO states but can update volumes within LOBs, allowing for order cancellations and updates, with minor changes to the current prototype. Recent LOB research shows promising directions for future exploration, such as parallel processing of multiple LOBs within the network~\cite{frey2023jax}. This work may inspire more industry-driven improvement of current programmable switches to support such extensions. 


\section{Conclusion}\label{ch7-Conclusion}

This paper presented a time-sensitive application of in-network ML for HFT. LOBIN, a prototype for constructing and updating limit order books within the data plane, was designed and deployed on programmable network devices. The evaluation shows that LOBIN can maintain high prediction accuracy compared with a server-based benchmark while achieving ultra-low microsecond-scale latency. More specifically, it provides over 10\% improvement in latency, compared with the NASDAQ order-matching server benchmark. The prediction performance of LOBIN can be further enhanced using a hybrid deployment. This approach proved highly effective, processing about 45\% of the traffic and 38\% of the potential transaction value directly on the switch, and keeping the average change in error rate at a stable 3\%. LOBIN's automated implementation streamlines the process of testing new stocks, using different data feeds, and experimenting with other ML models, marking a significant step towards true in-network trading applications.

\section*{Acknowledgment}

This work was partly funded by VMware and we acknowledge support from Intel. For the purpose of Open Access, the author has applied a CC BY public copyright license to any Author Accepted Manuscript (AAM) version arising from this submission.


\Urlmuskip=0mu plus 1mu\relax
\bibliographystyle{IEEEtran}
\bibliography{sigproc} 

\end{document}